\documentstyle[prl,aps,psfig]{revtex}
\newcommand{\be}{\begin{equation}}
\newcommand{\ee}{\end{equation}}
\newcommand{\ba}{\begin{eqnarray}}
\newcommand{\ea}{\end{eqnarray}}

\newcommand{\RR}{{\cal R}}
\begin{document}
\draft
%\twocolumn
\title{Electrically Charged Vortex Solution in Born-Infeld Theory}
\author{E.\ Moreno$^a$\thanks{Supported by CUNY Collaborative 
Incentive Grant 991999}\,,
C.~N\'u\~nez$^{b}$\thanks{CONICET}\,
and\,
F.A.\ Schaposnik$^b$\thanks{Investigador CICBA, Argentina}
\\
~\\
{\normalsize\it
$^a$ Physics Department, City College of the City University
of New York}\\
{\normalsize\it
New York NY 10031, USA}\\
{\normalsize\it
Physics Department, Baruch College, The City University of
New York}\\
{\normalsize\it
New York NY 10010, USA}
 \\
~\\
{\normalsize\it
$^b$Departamento de F\'\i sica, Universidad Nacional de La Plata}\\
{\normalsize\it
C.C. 67, 1900 La Plata, Argentina}}
\date{\today}
\maketitle
%===================================================================
\begin{abstract}
We obtain electrically charged vortex solutions for the 
Born-Infeld Higgs system with a Chern Simons term. We
analyse numerically these solutions, comparing their
properties with those of ``normal'' Nielsen-Olesen
vortices and also show that no charged vortex solutions exist
in Born-Infeld theory when the Chern Simons term is absent.
\end{abstract}


\bigskip

%===================================================================
%\newpage

\section{Introduction}
Born-Infeld (BI)
electrodynamics \cite{B}-\cite{BI} has recently received much attention in
connection with string theory and brane dynamics
\cite{Tse}-\cite{Lei} (see \cite{Pol} for a review and references). Classical
configurations have been studied for different models in
which the gauge field dynamics is governed by a BI Lagrangian and
in this way domain wall, vortex and monopole solutions were constructed
\cite{NS1}-\cite{G}
and also gravitating solutions relevant in supergravity
were found \cite{G}-\cite{GR}. Supersymmetry and BPS 
saturated solutions in connection with D-brane dynamics have also been
investigated \cite{CM}-\cite{GGT}.

An important point in the BI theory is that it admits finite energy
static solutions (originally proposed in \cite{B}-\cite{BI}
to describe the electron) which are not source free like solitons
but need pointlike sources which can be interpreted as
the ends of electric-flux carrying strings. In particular, for some
solutions discussed in \cite{G} (called there BIons)   strings
look like tubes joining smoothly onto a p-brane.

The solutions we find in the present work are not BIonic but source
free regular solitons of a BI theory. They are
related to Nielsen-Olesen
strings and in this sense they also look
like tubes which are in fact {\it electrically charged}
vortices (in contrast  with those presented in \cite{NS1} which
carry magnetic flux but no electric charge). 

Since in contrast with Maxwell electrodynamics  BI 
theory allows for \underline{finite}
 energy solutions which correspond to the
field of electric charges, 
one could envisage to construct 
finite energy electrically charged vortex solutions just
by studying a Born-Infeld-Higgs system, the non-linear electrodynamics
version of the Nielsen-Olesen model \cite{NO}
(which does not admit finite energy charged solutions). 
As we shall see, even
allowing singularities as electric sources for the vortex charge, such
solutions do not exist. Only if one adds a Chern-Simons (CS)
term, finite energy BI charged
vortices can be found, in analogy with
what happens in the Maxwell case
\cite{PK}-\cite{JW}.

Of course, introducing a CS term implies that 
static, $z$-independent vortices in $3+1$ dimensions
should be considered as static 
configurations in $2+1$. Hence,
they should be thought more
as planar configurations than as string-like
objects but one should expect
 that their main properties
should coincide with those arising when instead of a CS term 
(forcing to work in $2+1$) one
couples fermions to the BI model  \cite{JR}. It should be also noted that
the  resulting electric field   
is radial, in contrast with the configurations discussed in \cite{CM}
in which the direction of the electric field defines the orientation of the string.

The plan of the paper is the following: in section II we discuss the Born-Infeld-Chern-Simons  model coupled to a Higgs field. We show that 
charged vortex solutions exist in this model by solving numerically
the Euler-Lagrange equations. The relevant
properties of the solution are also
discussed in this section. We then show
in section III  that, as it happens for Maxwell
theory, Born-Infeld-Higgs theory (without CS term)
does not admit finite energy 
regular solutions corresponding to magnetic vortices carrying
electric charge. We give in section IV a brief summary of our results.

\section{The Born-Infeld-Chern-Simons-Higgs model}

We shall consider the $2+1$ dimensional
Lagrangian density for a complex scalar 
field $\phi$ minimally coupled to a $U(1)$ gauge field with dynamics
governed by a Born-Infeld (BI) action \cite{B}-\cite{BI} plus
a Chern-Simons action 
\be
L = 
-\beta^2 
\left( \sqrt{1 + \frac{1}{2\beta^2} F_{\mu\nu}F^{\mu \nu}} - 1 \right)
+ \frac{\kappa}{4\pi} \varepsilon^{\mu \nu \alpha} A_\mu F_{\nu \alpha}
+\frac{1}{2} D_\mu\phi^* D^\mu\phi  - V[\phi]
\label{1}
\ee
with
\be
F_{\mu\nu} = \partial_\mu A_\nu - \partial_\nu A_\mu ~,
\label{2}
\ee
\be
D_\mu \phi = (\partial_\mu + ie A_\mu)\phi
\label{3p}
\ee
and $V[\phi]$ a symmetry breaking potential such that
\be
\left.\frac{\delta V[\phi]}{\delta \phi}\right\vert_{\phi = \eta} = 0
\label{3}
\ee
The constant $\beta$ has the dimensions of the electromagnetic field,
$[\beta] = m^{3/2}$, while $[e] = m^{1/2}$ and $[\kappa] = m$. For
fields much weaker than the critical field value $\beta$, Lagrangian
(\ref{1}) reduces to the Maxwell-Chern-Simons-Higgs Lagrangian. 

It should be noted that models in which the kinetic term for
the Higgs field appears together with the $F_{\mu\nu}^2$ in the
square root can be considered. This possibility
was discussed
in \cite{NS1} in connection with Bogomol'nyi equations for BI systems.  
We will discuss this alternative in section III.

The equations of motion resulting from (\ref{1}) are
\be
\partial_\mu\left(\frac{F^{\mu \nu}}{\RR} \right)
+ \frac{\kappa}{2\pi} \varepsilon^{\mu \nu \alpha} F_{\nu \alpha}= 
 j^\nu
\label{4}
\ee
and
\be
 \Box \phi + 2ieA_\mu\partial^\mu\phi - e^2 A_\mu A^\mu \phi = - 
\frac{\partial V}{\partial \phi^*}
\label{5}
\ee
with
\be
\RR = \sqrt{1 + \frac{1}{2\beta^2} F_{\mu\nu}F^{\mu \nu}} 
\label{5a}
\ee
and
\be
j^\nu = \frac{ie}{2}(\phi^* \partial^\nu \phi - \phi \partial^\nu \phi^*) - 
e^2|\phi|^2 A^\nu
\label{6}
\ee
In Born-Infeld theories, the electric field 
$E^i$ and magnetic field $B^i$ are defined as usual as
\be
E^i = F^{0i} \hspace{3 cm}  B^i = \frac{1}{2}\varepsilon^{ijk}F_{jk}
\label{7}
\ee
while the electric induction $D^i$ and the magnetic intensity
$H^i$, take the form
\be
D^i = f^{0i} \hspace{3 cm}  H^i = \frac{1}{2}\varepsilon^{ijk}f_{jk}
\label{8}
\ee
with
\be
f_{ij} =\frac{ F_{ij}}{\RR}
\label{9}
\ee
In the limit $\beta^2 \to \infty$ the BI action coincides with the 
Maxwell action so that Lagrangian (\ref{1}) becomes
the usual Maxwell-Chern-Simons-Higgs Lagrangian
  and $D^i \to E^i$, $H^i \to B^i$.

 As it is well
known, the Abelian Higgs model with the usual Maxwell action 
(and no Chern-Simons term) has 
finite energy  axially symmetric
static solutions, the well-honored Nielsen-Olesen vortices \cite{NO}.
Vortex solutions carry quantized magnetic flux but,
in contrast with monopoles \cite{JZ}, they are necessarily
neutral since the existence of an electric charge implies infinite
energy. Indeed, consider the $\beta^2 \to \infty$ limit 
 in which Lagrangian (\ref{1}) becomes, for $\kappa = 0$,  
the usual Abelian Higgs  model
Lagrangian with an electric charge   $Q$
 which can be  defined 
as
\be
Q \equiv \int d^2 x j^0 =  \int d^2x \partial_i E^i = 
\int_{S^\infty} dx_i E^i
-  \int_{S^0} dx_i E^i
\label{10}
\ee
where
$S^\infty$ and $ S^0$ are the circles at $\rho\to \infty$
and $\rho \to 0$ respectively. Consistency of the 
coupled equations of motion
imply that $E^i \to 0$ at infinity so one needs $|\vec E| \to 1/\rho$
as $\rho \to 0$ in order to have non-zero electric charge. But this
behavior implies that the vortex has infinite energy.
The only way to evade this no-go result is to make dynamics
governed by something else than the Maxwell action.
In particular, one can add a Chern-Simons term endowing
the vortex with charge through its
well-known relation with magnetic flux
forced by eqs. of motion \cite{DJT}. This
has been worked out in refs. \cite{PK}-\cite{JW} where Abelian and non-Abelian
electrically charged vortices have been thoroughtfully investigated.

Another possibility is related to   the Born-Infeld electromagnetism
which allows   singular electric induction
$\vec D$   but with a regular electric field
$\vec E$ and a finite energy. One may expect
that regular charged vortex solutions could be constructed 
starting with a Born-Infeld
Lagrangian instead of a Maxwell Lagrangian, 
even in the absence of Chern-Simons terms. 
As we shall see, this is not the case and
the Born-Infeld-Higgs system also requires a Chern-Simons term in order
to exhibit finite energy solutions. We postpone
this issue   and first consider
the complete theory, including the Chern-Simons action.

In polar coordinates the adequate axially symmetric ansatz reads
\be
\phi(\vec x) = f(\rho) \exp(-in\varphi)
\hspace{2 cm} A_\varphi(\vec x) = -\frac{1}{\rho}A(\rho) \hspace{2cm}
A_0 (\vec x) = A_0(\rho)
\label{11}
\ee
We impose as boundary conditions
\be
 ~ \lim_{\rho \to \infty} f(\rho) = \eta \hspace{3.2cm} \lim_{\rho \to \infty} 
A(\rho) = -\frac{n}{e} \hspace{1.9cm} \lim_{\rho \to \infty} A_0(\rho) = 0
\label{12}
\ee
Concerning the origin
\be
f(0) = A(0) = 0 \hspace{2cm} A_0(0) = c
\label{13}
\ee
with $c$ a constant to be determined \cite{CLMS}.
With this conditions one can easily see that the magnetic flux
is quantized
\be
\Phi = \oint A_i dx^i = -\frac{2\pi}{e} n
\label{14}
\ee
Concerning the electric charge we use the formula (cf with (\ref{10}))
\be
Q = \int d^2 x j^0 =
 \frac{\kappa}{2\pi}\int d^2x\varepsilon^{ij}F_{ij} -\int d^2x \partial_i D^i 
=  \frac{\kappa}{2\pi} \Phi \left.
-2\pi \rho D_\rho\right\vert^\infty_0 
\label{15}
\ee
As already recognized in \cite{DJT}, the presence of the Chern-Simons term
forces the magnetic vortex to carry an electric charge.
There is also in the present case, a second possible contribution since
although the  $D_\rho$ field has a long distance damping effected by the
``photon'' mass, one could in principle have 
 a behavior at the origin of the form 
\be 
\label{17}
D_\rho(0) \sim -\frac{q}{2\pi}\frac{1}{ \rho} ~,  \hspace{0.5cm} r \sim 0
\ee
contributing to the electric charge, 
\be
Q = -\frac{\kappa}{e}n + q
\label{18}
\ee
The compatibility of
the imposed conditions at the origin and at infinity
should be investigated
through the analysis of the equations of motion (\ref{4})-(\ref{5}).
To this end,  it will be convenient to define dimensionless quantities
\ba
x(\tau) \! & = & \! n + eA(\tau) \nonumber\\
y(\tau) \! & = & \!  A_0(\tau)/\eta \nonumber\\
z(\tau) \! & = & \! f(\tau)/\eta
\label{20}
\ea
with 
\be
\tau = e \eta \rho
\label{21}
\ee
In terms of the new variables, eqs.(\ref{4})-(\ref{5}) become
\be
\tau \frac{d}{d\tau} \!\left( 
\frac{1}{\RR} \frac{\dot x}{\tau} 
\right) -  z^2 x= \delta \dot y \tau
\label{22}
\ee
\be
\frac{1}{\tau} \frac{d}{d\tau}\! \left( 
\frac{\tau}{\RR} {\dot y} \right)  - z^2 y = \delta \frac{\dot x}{\tau}
\label{23}
\ee
\be
\frac{1}{\tau} \frac{d}{d\tau}(\tau \dot z) + z(y^2 - 
\frac{x^2}{\tau^2}) - V'(z)   = 0
\label{24}
\ee
with
\be
\RR = 
\sqrt{1 + \frac{1}{\bar \beta^2}(\frac{\dot x^2}{\tau^2} - \dot y^2)
} 
\label{25}
\ee
\be
\bar \beta = \frac{\beta}{e \eta^2},\ \ \ \delta = \frac{\kappa}{\pi e \eta},\ \ \ 
V' = \frac{1}{e^2 \eta^4} \frac{\delta V}{\delta z}
\label{27}
\ee

We have solved this equations numerically, by minimizing 
the action associated with Lagrangian (\ref{1})
(Being the CS action metric independent, it does not
enter explicitly in the expression for the energy).   In fact, 
in terms of variables (\ref{20}) the
expressions for the action $S$ (omitting the time integral
since we are looking for static solutions) and the energy $E$ for our model
take the form
\be
S = 2\pi \eta^2 \bar \beta^2\int \tau d\tau 
\left(1 - R + \frac{1}{2 \bar \beta^2}(-\frac{x^2z^2}{\tau^2} - \dot z^2 + 
y^2 z^2) + \delta \frac{\dot x}{\tau} y -\frac{1}{\bar \beta^2}\frac{\lambda}{8e^2}(z^2 - 1)^2
\right)
\label{acc}
\ee
\be
E = \int d^2x T_{00} =  2\pi \eta^2 \bar \beta^2 \int \tau d\tau 
\left( \frac{1}{R} (1 + \frac{1}{\bar \beta^2} \frac{\dot x^2}{\tau^2})
- 1 + \frac{1}{2 \bar \beta^2} (\dot z^2 + y^2 z^2 + z^2 \frac{x^2}{\tau^2})
+\frac{1}{\bar \beta^2}\frac{\lambda}{8e^2}(z^2 - 1)^2
\right)
\label{ene}
\ee
Here $T_{\mu\nu}$ is the energy momentum tensor and we have used 
for the potential the usual form
\be
V[\phi] = \frac{\lambda}{8}(|\phi|^2 - \eta^2)^2
\label{pot}
\ee
As noted above, the CS coefficient $\delta$ enters in the action but not
in the energy of the vortex.

We shall now analyse the vortex equations 
(\ref{21})-(\ref{23}). For simplicity we will concentrate in the case $n=1$.
As a first step, let us study the asymptotic 
behavior of the solutions.
If one assumes that the boundary conditions at infinity, eq.(\ref{12}), are reached
exponentially, the differential equations for large $\tau$ 
can be linearized around $x$, $y$ and $1-z$. 
It is then straightforward to deduce the asymptotic behavior of the 
solutions, which is independent of the value of $\beta$:
\ba
x(\tau)&=&c_1\,{\tau} K_1(\alpha {\tau})\nonumber\\
y(\tau)&=&c_2\, K_0(\alpha {\tau})\nonumber\\
z(\tau)&=&1 - c_3\, K_0(\sqrt{\lambda} {\tau})\nonumber\\
\label{asym}
\ea
Here either
\be
c_1=c_2\ \ \ \ \ \ \ {\rm and}\ \ \ \ 
\alpha=\frac{\delta + \sqrt{\delta^2 +4}}{2}
\ee
or
\be
c_1=-c_2\ \ \ \ \ \ \ {\rm and}\ \ \ \ 
\alpha=\frac{-\delta + \sqrt{\delta^2 +4}}{2}.
\ee
It can be proved that for $\delta>0$ only the solution with 
$c_1=c_2$ exists while for $\delta<0$ only the solution  with 
$c_1=-c_2$ exists \cite{CLS}.

\begin{figure}%[b]
\centerline{
\psfig{figure=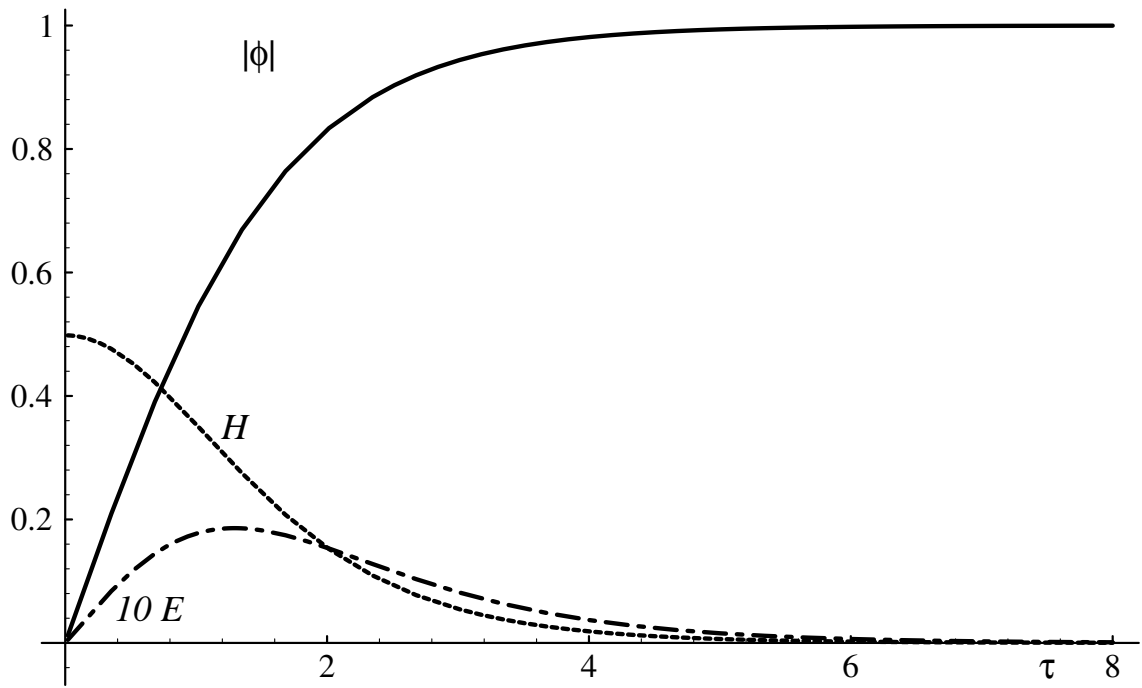,width=15cm,angle=0}}
\smallskip
\caption{
Vortex solution for $\beta=10$, $\delta=0.1$ and $\lambda=1$.
The full line corresponds to the modulus of the Higgs field,
$|\phi(\tau)|$, the dashed line correspond to the
magnetic field $H(\tau)$ and the dot-dashed line to the
electric field $E_r(\tau)$.
\label{beta10}
}
\end{figure}

\begin{figure}[t]
\centerline{
\psfig{figure=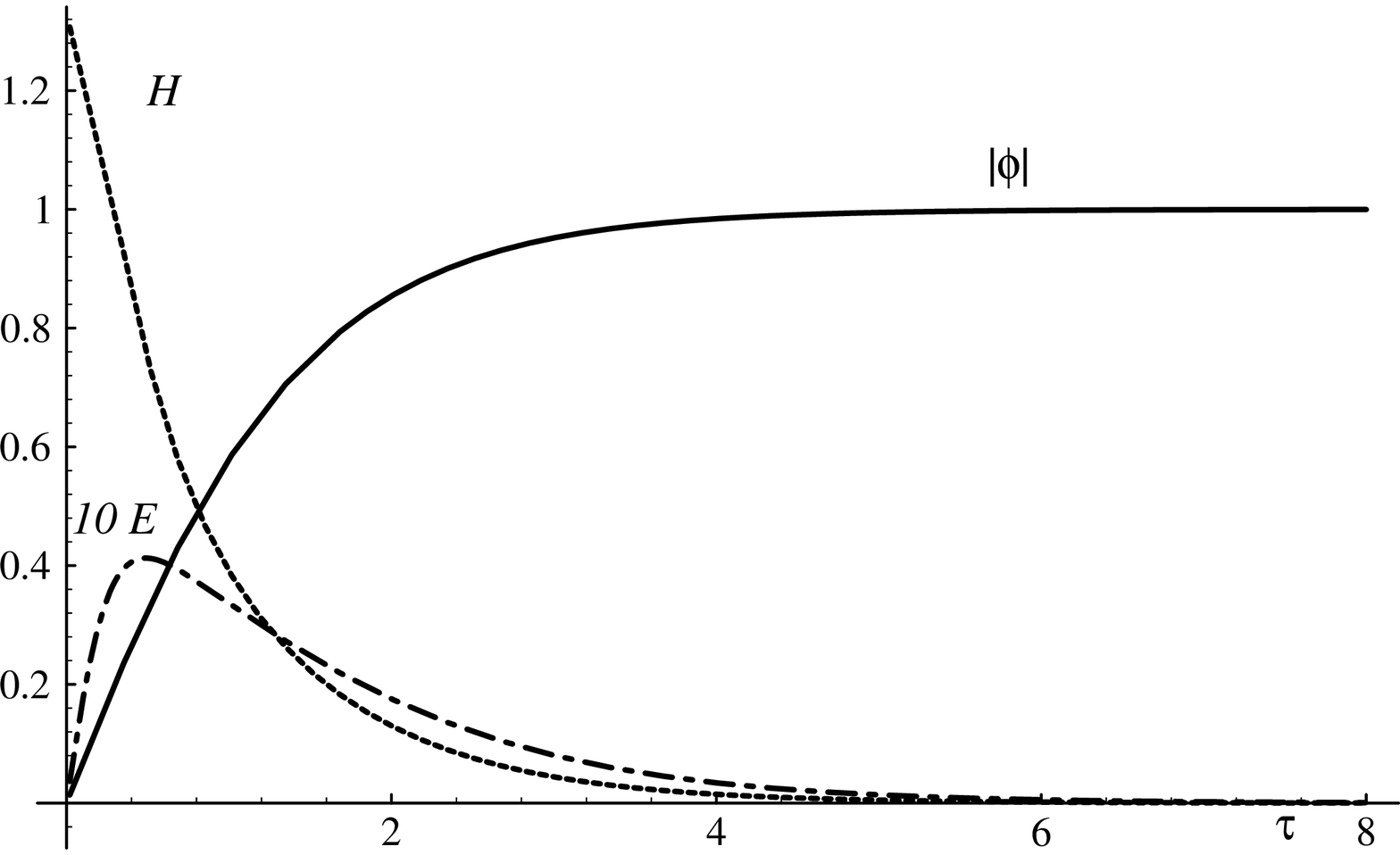,width=15cm,angle=0}}
\smallskip
\caption{
Vortex solution for $\beta=0.5$, $\delta=0.1$ and $\lambda=1$.
The full line corresponds to the modulus of the Higgs field,
$|\phi(\tau)|$, the dashed line correspond to the
magnetic field $H(\tau)$ and the dot-dashed line to the
electric field $E_r(\tau)$.
\label{beta05}
}
\end{figure}

To obtain a detailed profile of the vortex solution we solved
numerically the differential equations using a relaxation method 
for boundary values problem \cite{NR}. Such method 
determines the solution by starting with a guess and improving it
iteratively. 
For the case under consideration, the initial guess
\ba
x(\tau)&=&1-(1-e^{-a \tau})^2\nonumber\\
y(\tau)&=&c(1-(1-e^{-b \tau})^2)\nonumber\\
z(\tau)&=&1-e^{-d \tau}
\label{initial}
\ea
perfectly works with any reasonable election of the parameters 
$a,b,c$ and $d$. By minimizing the action with the ansatz (\ref{initial}),
(with $\beta=1$, $\delta=0$ and $\lambda=1$)
we choosed the values $a=0.67, b=0.78, c=-0.03,
d=0.86$.
The profile of the vortex solutions for different values of $\beta$,
for $\delta=0.1$ and $\lambda=1$ are shown in figs. \ref{beta10},\ref{beta05}.
Note that the solution we have found implies that the
behavior given in (\ref{17}) is to be excluded so that the
vortex electric charge is related to its 
magnetic flux in the usual form dictated
by the CS term, $Q = -(\kappa/e) n$.

\begin{figure}[b]
\centerline{
\psfig{figure=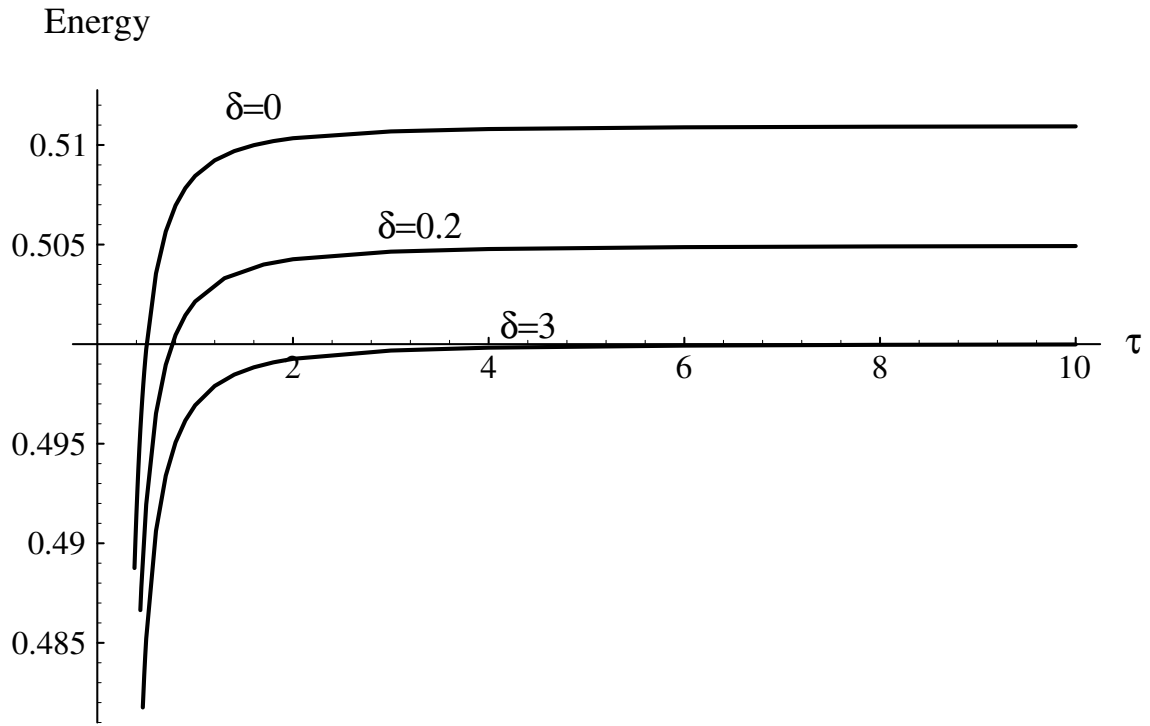,width=15cm,angle=0}}
\smallskip
\caption{
Energy of the vortex as a function of $\beta$ for different values 
of $\delta$ ( and $\lambda=1$).
The first line from the top corresponds to $\delta=0$ (non-charged vortex),
the second line corresponds to $\delta=0.2$ and the last one to
$\delta=3.0$.
\label{energy}
}
\end{figure}

It can be seen from the numerical solutions that, as expected,
for large values 
of $\beta$ the solution differs very little from the
Maxwell-Chern-Simons-Higgs vortices \cite{PK}\cite{CLS}. 
However as $\beta$ decreases, 
the vortex profile changes notably and a remarkable
peculiarity takes place: as $\beta$ approcahes to some critical value 
$\beta_c$, the magnetic field at the origin tends to infinity, and the 
numerical solution ceases to exist if $\beta$
is smaller than $\beta_c$ (see fig.~3)
This actual value of $\beta_c$ depends
on  the remaining parameters  ($\delta$ ,
$\lambda$) and its existence becomes more evident if
one represents 
the   vortex energy versus $\beta$ (for fixed values
of $\delta$ and $\lambda$, see fig. \ref{energy}). 
One can see in particular that
the derivative of the energy with respect to $\beta$ diverges as
$\beta$ approaches 
$\beta_c$. Even for $\delta=0$ (absence of Chern simons term) the 
$\beta_c$ is nonzero (and in fact takes its maximun value). 
Values of $\beta_c$ for differents values of $\delta$ are:
$\delta=0.465$  for $\beta_c=0$, $\delta=0.448$  for $\beta_c=0.1$, 
$\delta=0.416$  for $\beta_c=0.2, $ and $\delta=0.247$  for $\beta_c=1$.

Though we were unable to find an analytical argument to account 
for the
existence of this singularity, we have enough numerical evidence 
that 
supports our claim. In fact, in addition to the singular behavior 
of the energy 
around $\beta_c$,
and the absence of numerical solutions with $\beta<\beta_c$ using 
relaxation
methods, we found a similar pattern when solving the equations by
other methods (shooting algorithm for boundary value problem, etc.)

\section{ Non-existence of charged solutions in the absence of a CS term}
Let us first consider, instead of (\ref{1}), a Born-Infeld-Higgs Lagrangian
without the Chern-Simons term but including a $1/\beta^4$ term coupling
between the field strength and the Higgs field (appart from entering the
Higgs field kinetic term in the BI square root)
\be
L = \beta^2 \left(\sqrt{1 - \frac{1}{2\beta^2} F_{\mu\nu}F^{\mu \nu}
-\frac{1}{\beta^2}D^\mu \phi^* D_\mu \phi
+\frac{1}{\beta^4}{\tilde F}_{\mu} {\tilde F}_{\nu }
D^\mu \phi^*D^\nu \phi - \frac{1}{2\beta^4}
D_\mu \phi* D^\mu \phi F_{\alpha \beta}F^{\alpha \beta}
} -1 \right) - V[\phi]
\label{no5}
\ee
This Lagrangian is the $2+1$ Abelian version of the one exhibiting
dyonic solutions, investigated in
\cite{NS2}. In this last work it was shown that breaking of duality invariance
$ F_{\mu \nu} \to  {\tilde F}_{\mu \nu}$ 
was necessary in order to
have electrically charged monopoles, the coupling of the dual field
${\tilde F}_{\mu \nu}$ to the Higgs field doing this work. In the present 
$2+1$ dimensional case, one has to couple
the scalar to the dual $\tilde F_\mu = \frac{1}{2} 
\varepsilon_{\mu \nu \alpha} F^{\nu \alpha}$ and the simplest way is 
that given in (\ref{no5}).
 Moreover, inclusion of the Higgs
field kinetic energy term in the Born-Infeld square root
was necessary in order to have Bogomol'nyi equations. An ansatz of the
form (\ref{11}) together with the notation (\ref{20}) leads to the
following equations of motion for the gauge field components
\be
\tau \frac{d}{d\tau} 
\!\left[ 
\frac{1}{\RR} \frac{\dot x}{\tau}
\left(
-1 + \frac{1}{\bar \beta^2}
(\dot z^2 + \frac{x^2}{\tau^2}z^2 ) 
\right) - \frac{1}{\RR}\frac{1}{\bar \beta^2}
 z^2 \dot y y \frac{x}{\tau}
 \right] = \frac{z^2}{\RR} 
\left[ x - \frac{1}{\bar \beta^2}
\left( 
x\dot y^2 + \dot y \dot x y - x(\frac{\dot x^2}{\tau^2}- \dot y^2)
 \right) \right]
\label{n22}
\ee
\be
\frac{1}{\tau} \frac{d}{d\tau}\! \left[ 
\frac{\tau}{\RR} {\dot y} \left(1 + \frac{1}{\bar \beta^2}
(y^2 z^2 - \dot z^2 - \frac{x^2}{\tau^2}z^2)
 \right)
-\frac{1}{\RR}\frac{1}{\bar \beta^2} 
(\dot x y z^2 \frac{x}{\tau}
+\dot y \frac{x^2}{\tau} z^2 )
 \right] =
\frac{z^2}{\RR}\left(
-y + \frac{1}{\bar \beta ^2}(y \dot y^2 - \dot y \dot x \frac{x}{\tau^2}
\right)
 \label{n23}
\ee

We now assume a 
vortex-like behavior near the origin of the form
\be
x = 1 + \frac{1}{2} d_1 \tau^2
\label{29}
\ee
\be
y = c + y_1 \tau + \frac{1}{3} y_3 \tau^3
\label{30}
\ee
\be
z = a_0 \tau + \frac{1}{3}a_1 \tau ^3
\label{31}
\ee
The behavior (\ref{29}) corresponds to a vortex with $n=1$ units of magnetic
flux. This is consistent with the Higgs field having a simple zero at 
the origin as implied by (\ref{31}). The resulting magnetic and electric field
at the origin have the following behavior:
\be
\frac{\dot x}{\tau} = d_1
\label{32}
\ee
\be
\dot y = y_1 + y_3 \tau^2
\label{33}
\ee
Now, eqs.(\ref{n22})-(\ref{n23}) impose certain relations among the coefficients
in expansions (\ref{29})-(\ref{31}). Indeed, compatibility implies:
\be
a_0^2 =  \frac{\bar \beta^2}{2}
\label{35}
\ee
\be
c = 0 
 \label{36}
\ee
One should also note also that  
the behavior (\ref{17}) implies
\be
\frac{\dot y}{R} \sim \frac{q}{2\pi \eta} \frac{1}{\tau} \;\;\; \; {\rm for} \; \; \; \tau \sim 0  
\label{n34}
\ee
so that from (\ref{33}) we see that one
necessarily has $\RR \sim \tau$. Now, one can easily see
that formulae (\ref{35})-(\ref{36}) gives for $\RR$ a
behavior of the form
\be
\RR \sim \frac{a_0}{\bar \beta ^2}  \;\;\; \; {\rm for} \; \; \; \tau \sim 0  
\label{uni}
\ee
this showing that a charged vortex solution 
with finite energy does not exist in the theory with dynamics determined
by Lagrangian  (\ref{no5}). One can convince onself that other
possible $\beta^4$ terms do not change this situation.

\section{Summary}
We have studied the classical equations of motion
of a Born-Infeld Higgs model looking for (magnetic) vortex solutions
carrying also electric charge (the analogous of dyon solutions
in the case of monopoles). These kind of solutions do not exist
in Maxwell-Higgs model: electric charge implies 
in this case inifinite energy, as originally signaled in ref.\cite{JZ}.
Since precisely the problem of infinite energy of
charged solutions in pure electromagnetism
was overcome by Born and Infeld with
their proposal, one could expect that finite energy charged vortex could 
exist. We have proven however in section III that this is not the case.

On the contrary, 
finite energy charged vortices do exist when a Chern-Simons term is
added to the Maxwell Lagrangian. We have proven that this also happens
for the Born-Infeld Lagrangian. We have seen
numerically that for $\beta$ (the Born-Infeld parameter)
sufficiently large, the corresponding Born-Infeld
 solution differs very little from the Maxwell-Chern-Simons-Higgs vortices. 
When $\beta$ decreases the vortex profile changes notably and there
exists a critical value of $\beta$ where the magnetic field $H$ at the origin
and $\partial E/\partial \beta$
($E$ being the energy of the vortex solution) diverges. 
There is no solution for 
$\beta$ less then $\beta_c$. We have not
been able to find an analytical argument explaining this behavior
which we show is not a byproduct of our numerical procedure.

We conclude that vortex solutions in Born Infeld theories present
many interesting features which deserve a
thoroughfull study. In
particular, Bogomol'nyi equations should be analysed, not only for
their {\it per se}  interest but also in connection with
supersymmetric extensions and brane dynamics.
 We hope to report on these issues
in a forthcoming work.
 
~

\underline{Acknowledgements}: F.A.S. is
partially  suported
by CICBA, Argentina and a Commission of the European Communities
contract No:C11*-CT93-0315. This work is 
supported in part by funds provided by the U.S. Department of Energy (D.O.E.)
under cooperative research agreement \# DF-FC02-94ER40818.

%%%%%%%%%%%%%%%%%%%%%%%%%%%%%%%%%%%%%%%%%
% Figure 1
%%%%%%%%%%%%%%%%%%%%%%%%%%%%%%%%%%%%%%%%%

%%%%%%%%%%%%%%%%%%%%%%%%%%%%%%%%%%%%%%%%%
% Figure 2
%%%%%%%%%%%%%%%%%%%%%%%%%%%%%%%%%%%%%%%%%

%%%%%%%%%%%%%%%%%%%%%%%%%%%%%%%%%%%%%%%%%
% Figure 3
%%%%%%%%%%%%%%%%%%%%%%%%%%%%%%%%%%%%%%%%%

\end{document}